# Multi-directional Backlighting Compressive Light Field Displays

*Chen Gao, Sheng Xu, Yun Ye, Enguo Chen\**

Fujian Science & Technology Innovation Laboratory for Optoelectronic Information of China, Fuzhou, China

**\*Corresponding author: ceg@fzu.edu.cn**

**Abstract**

*We propose a compressive light field display of a wide viewing angle with a multi-directional backlight. Displayed layer images of sub-viewing zones are synchronized with the multi-directional backlight. Viewers can perceive a three-dimensional scene with a large viewing angle based on the persistence of vision.*

**Author Keywords**

compressive light field display; multi-directional backlight; wide viewing angle.

## 1. Introduction

Compressive light field displays can provide a high-resolution naked-eye 3D experience by utilizing the correlation between viewpoint images [1]. However, the viewing angle of the compressive light field display is narrow because the similarity of viewpoint images decreases as the reconstructed light field expands. Adding more display layers can improve the performance of compressive light field displays [2]. Nevertheless, the viewing angle cannot exceed 30°, even with six layers, because this method mainly expands the viewing range in the depth direction [3]. While the viewing angle can be adjusted to the viewers' positions by incorporating an eye-tracking device into the compressive light field displays, viewers would still experience the undesired crosstalk effect [4, 5]. Thus, increasing the angular bandwidth of the compressive light field displays is imperative.

Multi-directional backlights have been applied to expand the viewing angle of integral imaging 3D displays [6] and vector light field 3D displays [7]. The multi-directional backlight typically consists of a light-emitting diode (LED) array and a directional optics component. [8, 9]. Previous research on compressive light field displays introduces a time-multiplexed directional backlight as a low-resolution display layer to improve image quality [2]. In contrast, we argue that increasing the viewing angle of compressive light field displays is more urgent, as the image quality is relatively high within the narrow viewing zone. Inspired by recent work on light field displays with multi-directional backlights [6, 7], we propose a multi-directional backlighting compressive light field display of a wide viewing angle. The multi-directional backlight creates several sub-viewing zones with a small viewing angle, and a large viewing angle is reconstructed through these sub-viewing zones during the visual persistence period.

## 2. Principle

Figure 1 illustrates the principle of conventional compressive light field displays based on layered liquid crystal display (LCD) panels. A uniform backlight illuminates the diffuser of a certain scattering angle. Every ray of scattered light is modulated by the pixels of layered LCDs, creating a light field that contains rays of anisotropic intensity within the viewing zone. Theoretically, the maximum angle of the viewing zone is equal to the scattering angle of the diffuser, and the viewer will observe malposed layer patterns outside the viewing zone. However, as the viewing angle increases, the pixel workload also rises, leading to decreased image quality. Typically, the viewing angle of a compressive light field display with a uniform backlight and three layers is under 10°. To expand the viewing angle while preventing the workload of pixels from growing, we utilize the characteristic of the diffuser with a time-multiplexed multi-directional backlighting. Figure 2 illustrates the characteristics of the diffuser illuminated by light from normal and inclined incidence [10]. The incident beam is scattered in its propagation direction, and viewers would observe an image within the diffusion angle. In the persistence of vision, the perceived diffusion angle of normal and inclined incident beams is the superposition of the diffusion angles produced by each beam's passage through the diffuser. A seamless large viewing angle can be achieved by carefully adjusting the illumination beams' incident angles and the diffuser's scattering angle.

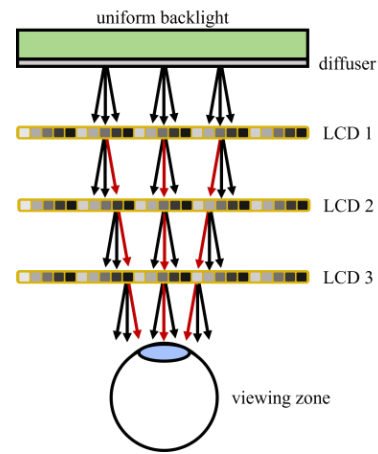

**Figure 1.** Principle of conventional compressive light field displays based on layered LCD panels.

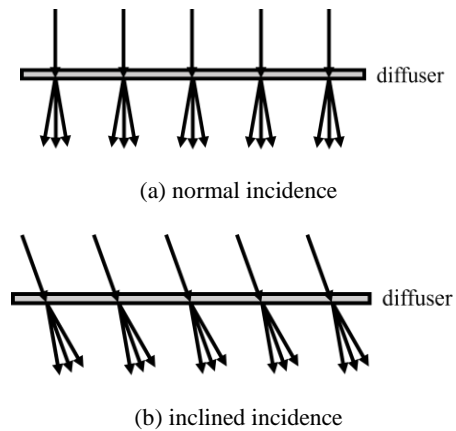

**Figure 2.** Light distribution when normal and inclined beams pass through the diffuser.

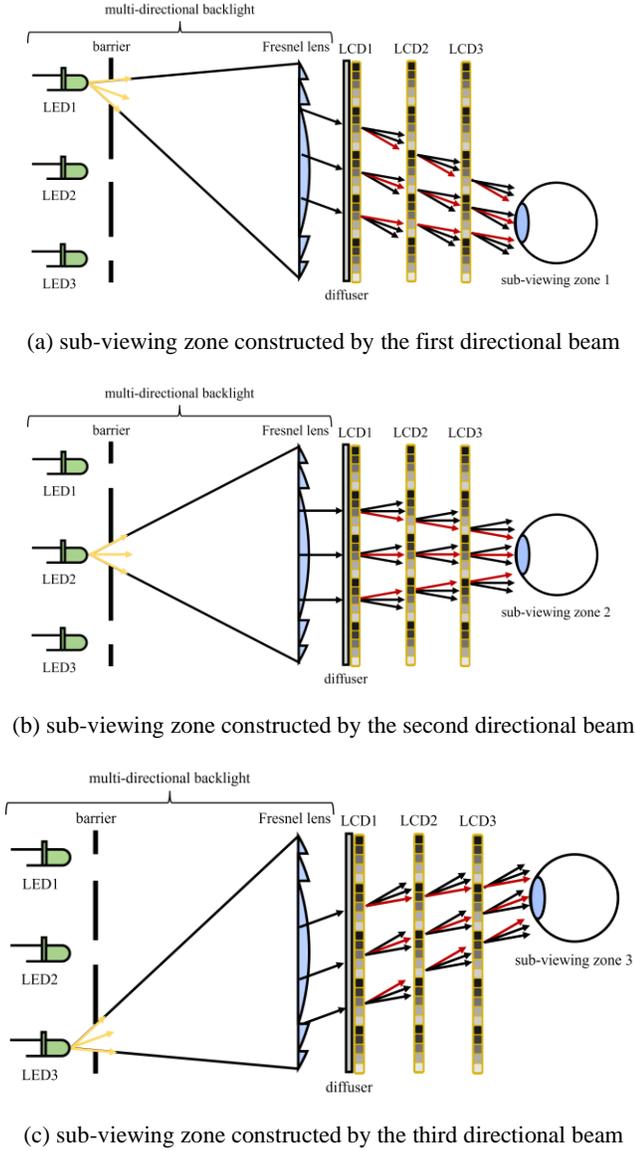

(a) sub-viewing zone constructed by the first directional beam

(b) sub-viewing zone constructed by the second directional beam

(c) sub-viewing zone constructed by the third directional beam

**Figure 3.** Principle of multi-directional backlighting compressive light field displays.

Figure 3 illustrates the principle of the multi-directional backlighting compressive light field display, which expands the viewing angle in one direction. The display system includes a multi-directional backlight, a diffuser, and layered LCD panels with a high refresh rate. A Fresnel lens, barrier, and LED array constitute the multi-directional backlight for simplicity. Since every LED's divergence angle is identical and symmetric, a barrier is needed to make outgoing light asymmetric. Other multi-directional components, such as lens array or stacked light guide plates, can replace the Fresnel lens. The advantage of light guide-based backlight is its high efficiency. Every LED source corresponds to a different direction of collimated illumination. When the LEDs in specific positions illuminate, a collimated light passes through the diffuser. Meanwhile, the layer patterns are uploaded to LCDs. The diffused light illuminates layered LCDs and reconstructs a light field within a sub-viewing zone of a small viewing angle. All LEDs are lit up sequentially within the flicker fusion threshold, and the viewer will perceive a spliced light field of a large angle. To satisfy the condition of flicker fusion, the refresh rate of LEDs and LCD panels should be at least 30 Hz times the number of sub-viewing zones. For example, LEDs and LCD panels with 90 Hz are needed to implement a multi-directional backlighting compressive light field display of 3 sub-viewing zones.

## 3. Lens and Barrier-based Multi-directional Backlighting Design Parameters

Figure 4 illustrates the calculation model for lens and barrier-based multi-directional backlight along the horizontal direction, including the single Fresnel lens backlight and the lens array backlight. A lens is illuminated by $M$ LEDs arranged in the $x$ direction, and $M$ is an odd integer. Let $D$ and $f$ be the aperture and focal length of the lens. The $i^{th}$ LED's position along the $x$ direction $x_i$ could be calculated by Equation 1, where $\theta$ is the diffusion angle of the diffuser.

$$x_i = f \tan[(\frac{M}{2} - i)\theta] \qquad (1)$$

The distance from the LED array to the barrier should satisfy Equation 2, where $\varphi$ is the LED's divergence angle.

$$d \leq \min_i(\frac{x_i}{2\tan\varphi}) \qquad (2)$$

Equation 3 expresses the length of holes in the barrier, where $\alpha_i$ and $\beta_i$ follow Equation 4.

$$h_i = d(\tan\alpha_i + \tan\beta_i) \qquad (3)$$

$$\begin{cases} \tan\alpha_i = (\frac{D}{2} - x_i)/f \\ \tan\beta_i = (\frac{D}{2} + x_i)/f \end{cases} \qquad (4)$$

The above analysis can be extended to a multi-directional backlight that broadens both horizontal and vertical viewing angles by incorporating LEDs and a barrier in the $y$ direction.

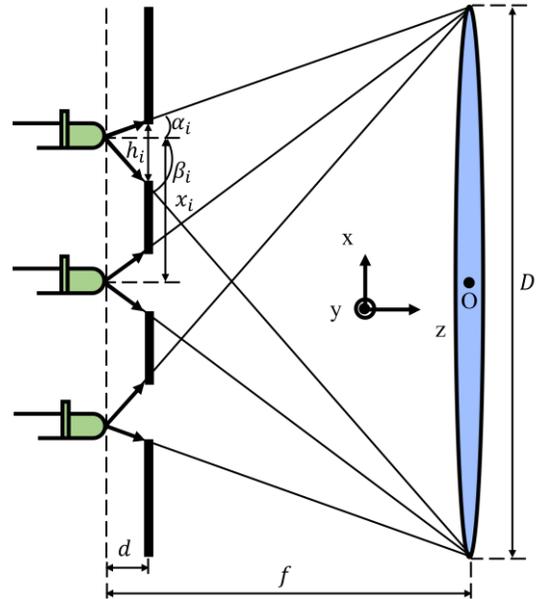

**Figure 4.** Calculation model of lens-based multi-directional backlight.

## 4. Simulation and Results

We simulated a multi-directional backlighting compressive light field display of a 30° horizontal viewing angle. The viewing zone is composed of 3 sub-viewing zones. Each sub-viewing zone is reconstructed by three layer patterns iterated from 7×7 parallax images within the viewing angle of 10°×10°. The layer spacing is equally set at 10 mm. To reduce backlight loss, the compressive light field display model of polarized layers is implemented [2]. Figure 5 shows displayed layer patterns of three zones, which are iterated with the Simultaneous Algebra Reconstruction Technique (SART) of 5 iterations. The time-sequential light field of a 30° viewing angle reconstructed by these layer patterns is illustrated in Figure 6. The reconstructed light field preserves a relatively high peak signal-to-noise ratio (PSNR) compared with the target light field. A software written in the OpenGL Shading Language (GLSL) implements factorizations of layer patterns and simulation results. The viewing angle of the multi-backlighting compressive light field could be further expanded if LEDs and LCD panels with higher refresh rates are available.

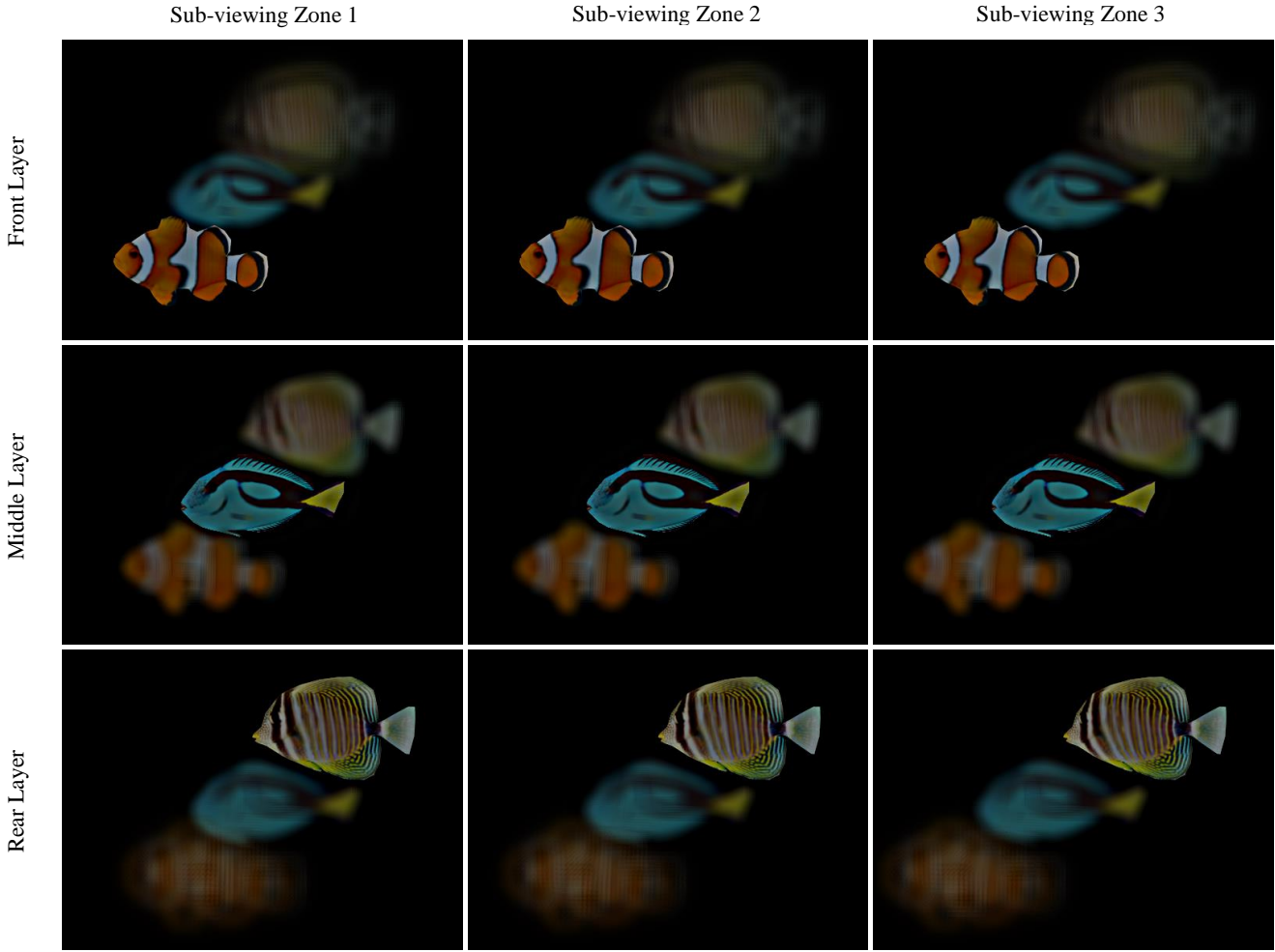

**Figure 5.** Displayed layer patterns of three sub-viewing zones.

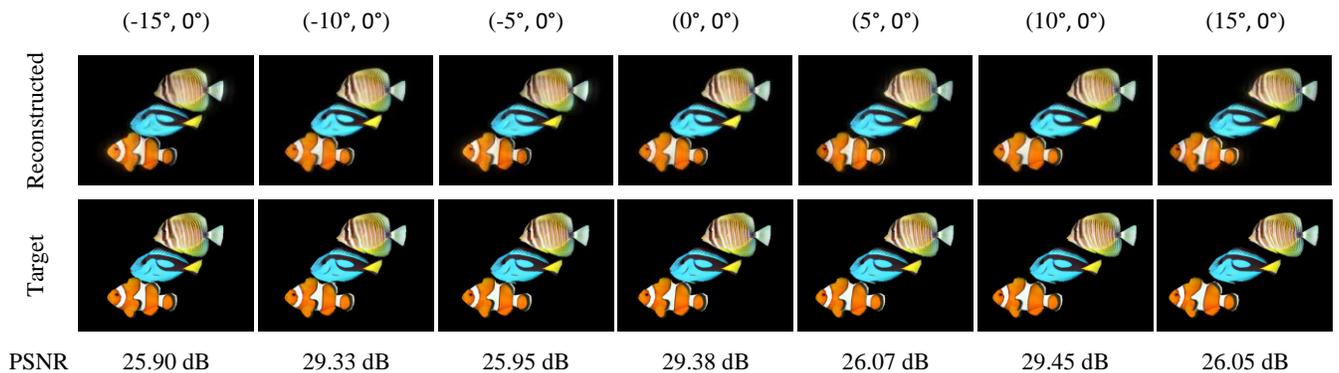

**Figure 6.** Visual and numerical comparison between reconstructed and target light field in simulation.

## 5. Conclusion

We have shown the feasibility of expanding the viewing angle of the compressive light field displays with a multi-directional backlight system. While the current simulated result only considers an ideal multi-directional backlight, we are building an experimental prototype of multi-directional backlighting compressive light field displays.

## 6. Acknowledgements

This work is supported by the Natural Science Foundation for Distinguished Young Scholars of Fujian Province (No. 2024J010046); Key Science and Technology Project Program of Fujian Province (No. 2024HZ022005); Fujian Province Technology Innovation Key Research and Industrialization Program (2024G020) and the Mindu Innovation Laboratory Special Program for Development of Industry-University-Research Integration (No. 2024CXY106).